\documentstyle[12pt]{article}
\begin{document}
\begin{titlepage}
\begin{center}
\vspace{72pt}
{\large \bf
DYNAMICS OF A DISORIENTED CHIRAL CONDENSATE    }
\end{center}
\vspace{36pt}
\begin{center}
{\sl Jean-Paul BLAIZOT$\, ^a$\footnote{Electronic mail:
blaizot@amoco.saclay.cea.fr} and Andr\'{e}
KRZYWICKI$\, ^b$\footnote{Electronic mail:
krz@qcd.th.u-psud.fr}}\\
\vspace{10pt}
$^a$Service de Physique Th\'{e}orique,
Centre d'Etudes de Saclay,      \\
F-91191 Gif-sur-Yvette, France\\
\vspace{5pt}
$^b$Laboratoire de Physique Th\'{e}orique
et Hautes Energies, B\^{a}t. 211,\\
Universit\'{e} de Paris-Sud, F-91405 Orsay,
France\footnote{Laboratoire
asoci\'{e} au C.N.R.S.}
\vspace{36pt}
\end{center}
\begin{center}
{\bf Abstract}
\end{center}

We use the linear $\sigma$ model to analyse
the dynamics of a disoriented chiral condensate.
For idealized boundary conditions
appropriate to high energy collisions, the
problem can be reduced to a one
dimensional one. The evolution of the chiral
state is then that of a
simple dynamical system and can be
studied analytically.

PCAC number(s): 25.75.+r, 13.85.Hd , 03.50.Kk
\vfill \par\noindent
February  1993\\
LPTHE Orsay 94/18 \\
SPhT T94/019 \\
\vspace{5pt}

\end{titlepage}

\newpage

\section{ INTRODUCTION }
The aim of this paper is to present a simple,
analytical discussion of the time
evolution of a disoriented chiral
condensate (DCC) which could be produced in a
high-energy collision \cite{ans}-\cite{bk}.
We shall extend our previous
investigation \cite{bk} and make contact with
recent works which have been
triggered by the idea of a ``quench",
put forward in ref. \cite{rw}. Specifically,
we shall study the classical equations
of motion of the linear $\sigma$ model,
assuming random initial
conditions for the fields and their
derivatives. This randomness could result from a
sudden cooling of a hot quark-gluon plasma,
as proposed in \cite{rw}, but we
shall not insist on an interpretation
based on initial thermal equilibrium.
The evolution of the system will
be described by the classical equations of
motion of the $\sigma$ model. We
neglect possible quantum effects
(for a recent discussion of
the decoherence of the radiation
off colliding heavy ions see \cite{krz}).
The main motivation for using
the linear, instead of the
non-linear $\sigma$ model as in \cite{bk},
is to allow for the possibility
to start the evolution from a
chirally symmetric random configuration, as in \cite{rw}.
As we shall see, however, many
features of our previous study survive.
\par
 As in ref.
\cite{bk} we adopt the idealization
originaly proposed by Heisenberg
\cite{heis}: we assume that at time
$t = 0$ the whole energy of the collision
is localized within an infinitesimally
thin slab with infinite transverse
extent. The symmetry of the problem
then implies that the fields depend
only on the invariant $\tau = \sqrt{t^2
- x^2}$, where $x$ is the longitudinal
coordinate. This is of course a
 dramatic simplification of the actual
situation. In particular, our
assumptions lead to unrealistic
correlations both in the
transverse plane and in rapidity.
However, with this
idealization the problem becomes
1+1 dimensional, and can be
treated analytically. This will allow
us to identify
clearly various regimes of the
dynamics of chiral
condensates in the presence of
the longitudinal
expansion which necessarily occurs
in case of high energy
collisions. We believe that such
analytical studies are
complementary to numerical simulations
\cite{rw,ggp,zhxnw}
aiming at getting a more realistic
picture, and they may prove useful in
understanding the results obtained in
such simulations.
\par
Although we use Heisenberg's
idealization to justify working
in 1+1 dimensions, we
restrict our discussion to proper
times $\tau > \tau_0$. An
extension of the model down to
$\tau = 0$ leads to
unphysical results and is
completely unjustified.
Heisenberg \cite{heis} has
pointed out that in a wide class
of non-linear theories with
gradient interactions, the fields
have to vanish like $\tau$ in
the limit $\tau \to 0$. It
seems more likely, however,
that the fields
($\vec{\pi},\sigma$) do not
tend to any well defined limit
as $\tau \to 0$ but behave
chaotically (in the loose sense
of the word). In any case, as $\tau
\to 0$ , the relevant degrees of
freedom are presumably no
longer hadronic and, besides,
the effective theory is
certainly more complicated
than the linear $\sigma$ model.
\par
Thus, we shall assume that
 starting from some finite time
$\tau_0$, we can describe the state of matter
produced in a collision by the
linear $\sigma$ model. Our
results will turn out to be little
 dependent of the
specific value of this time $\tau_0$,
 as long as it is chosen so
 that at $\tau_0$ the system is
in the chirally symmetric configuration. The
values
of the fields at this time are taken
as random variables.
In fact, as we shall show in section 2,
 the equations of
motion are, for short times,
equivalent to the conservation
of the isospin vector and axial
vector currents. The bulk of the
dynamics will be entirely
determined by the values of
these currents at $\tau_0$. In
section 3, we shall
identify several regimes of the
evolution of our chiral
condensate. We shall see that
the problem
reduces to that of finding the motion
of a fictitious particle in a time dependent
potential and subject to a time
dependent friction. The paper ends in
section 4 by a discussion.

\section{FORMULATION OF THE MODEL}
Our system is described by the Lagrangian
of the linear $\sigma$ model,
with the standard chiral symmetry breaking term:

\begin{equation}
L = {1 \over 2} [ (\partial\sigma)^2 +
(\partial\vec{\pi})^2] - {\lambda
\over 4}
(\sigma^2 + \vec{\pi}^2 - 1)^2 + H\sigma. \label{Lag}
\end{equation}

\noindent
The fields $\sigma$ and $\vec{\pi}$ are
dimensionless and the parameters $\lambda$
and $H$ have the dimension of [mass]$^{2}$.
We have scaled out a dimensionless factor $Sf_\pi^2$,
which would normally occur in the expression
of the Lagrangian for the 3-dimensional model
\cite{bk}. The parameters of the model may be related to the
masses of the $\sigma$ and pion field, respectively:
$m_\sigma^2 \approx 2\lambda$ and
 $m_\pi^2=H$. For the sake of the argument
we shall proceed as if one had $m_\sigma
\gg m_\pi$. We shall return to this point in the discussion.
\par
The equations of motion read

\begin{equation}
\tau^{-1} (\tau \sigma ' ) '  = - \lambda (\sigma^2 +
\pi^2 -1) \sigma + H,
\label{eqm1}
\end{equation}

\noindent
and

\begin{equation}
\tau^{-1} (\tau \vec{\pi}\; ') ' =
 - \lambda (\sigma^2 + \pi^2 -1 ) \vec{\pi},
\label{eqm2}
\end{equation}

\noindent
where the prime denotes differentiation
with respect to $\tau$. From
(\ref{eqm2}) one easily gets

\begin{equation}
\vec{\pi} \times \vec{\pi}\; ' = {\vec{a}
\over \tau} , \label{vcons}
\end{equation}

\noindent
where $\vec{a}$ is an integration constant. This is
a direct consequence of the conservation
 of the isovector current together
with the assumption that the fields depend
on $\tau$ only. Hence the component
of $\vec{\pi}$ along
$\vec{a}$ vanishes:

\begin{equation}
\pi_a = 0.
\label{1}
\end{equation}

\noindent
A further consequence of the equations
 of motion is the equation

\begin{equation}
\tau^{-1} [\tau ( \vec{\pi} \sigma ' -
\sigma \vec{\pi}\;')] ' = H
\vec{\pi} ,
\label{acons0}
\end{equation}

\noindent
which is just the statement of partial
conservation of the isoaxial-vector
current (PCAC).

For small enough $\tau$, the effect
of the symmetry breaking term
$H\vec\pi$
can be neglected. The isoaxial-vector current is then
conserved, and one can write, in analogy to (\ref{vcons})

\begin{equation}
\vec{\pi} \sigma' - \sigma \vec{\pi}\; '
= {\vec{b} \over \tau},
\label{acons}
\end{equation}

\noindent
where $\vec{b}$ is another integration constant.
When eqs. (\ref{vcons}) and
(\ref{acons}) are satisfied, the component
 of $\vec{\pi}$ along $\vec{c} =
\vec{a} \times \vec{b}$ equals $\sigma$,
up to the constant factor $a/b$.  The
forms of the conservation
laws (\ref{vcons}) and
(\ref{acons}) reveal important features
of our model. Because of the symmetry
of the fields, and the classical dynamics,
in the limit of vanishing pion
mass the isospin vector and axial vector
currents keep fixed orientations in
isospace as the system evolves.
Of course, any  orientation is equally
probable. But once fixed in the initial
conditions, it remains constant.
\par
The equations (\ref{vcons}) and
(\ref{acons}) are identical to those
obtained in the non-linear $\sigma$ model
\cite{bk}, and  they are independent
of $\lambda$. The non-linear
$\sigma$ model is obtained in the limit
$\lambda\to\infty$. In  this limit,
the motion takes place on the
hypersphere $\sigma^2+\vec{\pi}^2=1$ and is
entirely determined by the conservation
laws for the isospin currents. When
$\lambda$ is large, but not infinite, one
can still separate the general motion into
a motion on the sphere, which
one may loosely associate with the pion
degrees of freedom, and a motion
normal to the sphere which involves
the $\sigma$ degree of freedom. As long as
the parameter $H$ can be neglected, there
is no scale controlling the motion on
the sphere, while $\lambda^{-1/2}$ controls
the time dependence of the oscillations
normal to the sphere. When $H$ can no
longer be ignored, it controls the
time behaviour of pionic excitations.
\par
In order to analyse the evolution of
the chiral state, we shall use the
following parametrisation which takes
into account the symmetries of the
problem:

\begin{eqnarray}
\pi_b	& = & r \sin{\theta} \nonumber \\
\pi_c	& = & r \cos{\theta} \sin{\omega} \nonumber \\
\sigma & = & r \cos{\theta} \cos{\omega}.
\label{par}
\end{eqnarray}

Note that $\sigma^2+\pi^2=r^2$.
Our problem has now been
reduced to  a simple dynamical problem
for a point particle with one radial
degree of freedom, $r$, correesponding
to the $\sigma$
excitations, and two angular degrees of
freedom $\theta$ and $\omega$,
corresponding to the pionic excitations.
As will become clear later, the
evolution of the system depends upon
the initial conditions mostly via the
constants $a$ and $b$ (while the detailed initial
values of $r, \theta$, $\omega$
and their derivatives are not
essential). The two constants
$a$ and $ b$ determine the strength of
the isospin vector and axial vector
currents (note that due to a different
choice of variables, the present
 $a$ and $b$ are twice as large as those
introduced in ref.\cite{bk}). They
 are random variables whose
probability distribution will be calculated in the last section.
\par
As a final remark in this section,
we note that when eqs.
(\ref{vcons}) and (\ref{acons}) are
satisfied, the angle $\omega$ is
constant and given by

\begin{equation}
\omega = \arctan{{a \over b}}.
\label{asi}
\end{equation}
\noindent
Thus the motion is planar in the 3-space  ($\sigma, \pi_b,
\pi_c$). This plane becomes almost identical
to  $(\pi_b,\sigma)$ when $b \gg a$.
We shall find convenient, at some point in the
forthcoming discussion, to work in this particular limit.

\section{DCC AS A SIMPLE DYNAMICAL SYSTEM}

\noindent
 The initial time $\tau_0$ is supposed to be such
that the mass term $\propto H$ can be
 ignored at the beginning of the
evolution. Then the initial motion
is governed by the two
conservation laws for the isospin
currents, i.e. eqs. (\ref{vcons}) and
(\ref{acons}). These translate into the
following equation for  the ``angular
momentum'' of the fictitious point
particle associated with our system:

\begin{equation} r^2 \theta' =
{\kappa \over \tau}, \label{rot}
\end{equation}

\noindent
The equation  for the radial motion
is obtained by neglecting $H$ in the
equations of motion (\ref{eqm1})-(\ref{eqm2}),
multiplying the equation for
$\sigma$ ($\pi_b$) by $\cos{\theta}$
($\sin{\theta}$), and adding the
resulting equations. We get

\begin{equation}
r'' + {1 \over \tau} r' =
{{\kappa^2} \over {\tau^2 r^3}}
- \lambda r (r^2 - 1).
\label{req}
\end{equation}

\noindent
We have used eq. (\ref{rot}) to eliminate $\theta'$,
and also the fact that $\omega$
is constant when the isospin currents
are strictly conserved. The
second term on the left-hand side
may be interpreted as a time dependent friction.
Thus, the fictitious particle experiences damped
oscillations. The amplitude of these
oscillations can be initiallly quite
 important if $r(\tau_0)$ is small.
In that case, the repulsive force represented by the
first term on the right-hand side of (\ref{req})
provides to the particle a large
velocity. However,
the oscillations are rapidly
damped, while the repulsive force
itself decays fast with time.
\par
Further insight on the radial motion
can be gained by considering
the  time variation of the mechanical
energy of the fictitious  particle.
This is given by

\begin{equation}
\frac{d}{d\tau}\left[\frac{(r^\prime)^2}{2}+
\frac{\lambda}{4}(r^2-1)^2+\frac{\kappa^2}{2\tau^2r^2}\right]
=-\frac{\kappa^2}{r^2\tau^3}-
\frac{(r^\prime)^2}{\tau} < 0 .
\label{enr}
\end{equation}

\noindent
This derivative is
particularly large, in absolute value,  during the
very early stage of the evolution due to the
presence of the term $\sim \tau^{-3}$.
The whole damping
is of kinematical origin; it reflects the
decrease of the field energy
density as the system expands.
One observes that the influence of the
angular motion on the radial motion
quickly disappears. The energy in the
angular motion is controlled by $\kappa$.
It becomes of the same order of
magnitude as the potential energy
$\propto\lambda$
when $\tau\sim\kappa/\sqrt{2\lambda}$,
and is negligible at later times.
\par
For not too
small $\tau$ it is certainly a good
approximation to linearize the
expression appearing on the right-hand
 side of (\ref{req}). There results an
inhomogeneous Bessel equation for
 $r-1$. The solution takes a  particularly
suggestive form for $\sqrt{2 \lambda} \tau \gg \max[{\kappa^2,1}]$:

\begin{equation}
r \approx 1 + {C \over {(\sqrt{2\lambda}\tau)}^{1
 \over 2}} \cos{(\sqrt{2
\lambda} \tau + \eta)} ,
\label{r}
\end{equation}

\noindent
where $C$ and $\eta$ are integration
constants fixed by the initial
conditions. Thus, because of the
damping caused by the expansion, the
system is driven towards
the orbit $r = 1$ [strictly speaking, for
$H \neq 0$, the orbit would
rather be at $r = 1 + O(H/\lambda)$].
\par
When $r \approx 1$, the
integration of eq. (\ref{rot}) yields $\theta \approx
\kappa \ln{(\tau/\tau_0)}$
and one recovers the solution of the
non-linear $\sigma$ model
presented in ref. \cite{bk}. In
fact the approach to the orbit
is fairly slow, and a change of
regime will in most cases take place
before the orbit is actually reached.
The oscillations in the radial
motion have a large frequency,
$\propto\sqrt{2\lambda}$. Since the angular
motion, controlled by $\kappa$,
is generically slower than the radial one, one can,
with a good accuracy, replace $r^2$
in eq. (\ref{rot}) by its expectation value
over a few periods of the radial
 motion. This expectation value reaches
a value $\approx 1$ in a short time.
Therefore, the result $\theta \approx
\kappa \ln{(\tau/\tau_0)}$ can be taken
 as a good approximation  from
nearly the beginning of the evolution.
\par
A change of regime is expected to occur
 when the mass term $\propto H$ can
no longer be ignored. Let us, therefore,
 discuss in more detail the quality of the
approximation $H = 0$. Integrating eq.
 (\ref{acons0}) for $\pi_b$ one has

\begin{equation}
\tau (\pi_b \sigma' - \sigma \pi_b') =
b + H \int^{\tau} \tau r \sin{\theta} \; d\tau.
\label{Hb}
\end{equation}

\noindent
In estimating the integral
 above it is important to notice
that the integrand is oscillating, and
it does so quite rapidly when $\kappa$
 is large. In order to find a
 rough estimate, we replace again $r$ by
 1, its average value, so that $\theta
= \kappa \ln{(\tau/\tau_0)}$ (we assume
 $\theta(\tau_0)=0$ to avoid the
proliferation of inessential integration constants).
The integral in (\ref{Hb}) is then easily
 evaluated. For large $\kappa$, it
equals $(\tau^2/\kappa)
\cos{(\kappa\ln{(\tau/\tau_0)}}$. Therefore, the
neglect of $H$ is justified as long as

\begin{equation}
\tau^2\ll {{b\kappa} \over H}.
\label{reg1}
\end{equation}

\noindent
It remains to
determine the conditions ensuring that
$\omega =$ const., because the
 planarity of the motion in the 3-space
($\sigma,\pi_b, \pi_c$)
is another characteristic
feature of the early, chirally symmetric regime.
Using the polar parametrisation
(\ref{par}) and the equations (\ref{acons0})
and (\ref{vcons}) one finds
after some algebra

\begin{equation}
\cos{\theta} \sin{(\omega - \omega_0)}
= {H \over \kappa} [\cos{\theta} \sin{\omega}
\int^\tau \tau r \sin{\theta} d\tau -
 \sin{\theta} \int^\tau \tau r
\cos{\theta} \sin{\omega} d\tau]
\label{reg2}
\end{equation}

\noindent
where $\omega_0 = \arctan{(a/b)}$, the
value $\omega$ takes when $H = 0$. On
the right-hand side set $\omega =
\omega_0 +\delta\omega$ and expand
neglecting terms of O[$(\delta\omega)^2$].
Furthermore, set again $r=1$ and
$\theta =\kappa \ln{(\tau/\tau_0)}$ to get

\begin{equation}
\cos{\theta} \; \delta\omega \approx -
{{H\tau^2} \over {\kappa^3}} [a -
b \cos^2{\theta}] - {{bH} \over
{\kappa^2}} \sin{\theta} \int^\tau
\tau \cos{\theta} \delta\omega d\tau
\label{reg3}
\end{equation}

\noindent
One can see that, as long as $\cos{\theta}$
is not small, $| \delta\omega|$
is of order $H\tau^2/\kappa^2$.
However, when
$\cos{\theta} = 0$, the left-hand side
of (\ref{reg3}) vanishes
identically, and a perturbative (in $H$) calculation
of $\delta\omega$ is impossible. However, one easily sees that the
right-hand side of (\ref{reg3}) vanishes
when the time average $\langle \delta\omega
\rangle\approx
-a/b$. We
conclude that the angle $\omega$ oscillates with
an amplitude of order $a/b$, i.e. independent of $H$. We shall
assume herefrom that

\begin{equation}
b \gg a
\label{reg4}
\end{equation}

\noindent
The constraint (\ref{reg4}) makes the
problem technically simpler, since
$\omega$ remains small and the motion
on the orbit is quasi-planar at any time.
We do not think
that limiting the discussion to this
case is a severe restriction. The
general case is just more
cumbersome, without being
really more instructive. With (\ref{reg4})
we gain considerably in clarity.
Furthermore, only for large enough $b$
the problem is really interesting, a
point that will become more and more
clear as the discussion will develop.
Finally, as will be shown in the last
section, it is improbable that both $b$
and $a$ are large. When (\ref{reg4})
holds, eq. (\ref{reg1}) can be rewritten
as follows

 \begin{equation} \tau  \ll {b\over\sqrt{H}}.
\label{reg5}
\end{equation}

The presence of $H$ in the equations
of motion is an inessential complication
as long as (\ref{reg5}) is true. We
shall see  that a dramatic
change of regime occurs when
$\tau \approx b/\sqrt{H}$.
\par
Consider first what happens at large
times. To this end we use eqs.
(\ref{vcons}) and (\ref{acons0}).
 Working in the polar parametrization
(\ref{par}), with $r = 1$, one easily
convinces oneself that $\theta, \omega
 \to 0$ like some inverse power of
$\tau$ when
$\tau \to \infty$, modulo possible oscillations.
For small enough $\theta, \omega$ eqs.
(\ref{acons0}) read

\begin{eqnarray}
\tau^{-1} (\tau \theta')'  & = &
-H \theta \nonumber \\
\tau^{-1} (\tau \omega')' & = &
- H \omega
,\label{acons3}
\end{eqnarray}

\noindent
while eq. (\ref{vcons}) becomes a
constraint on the Wronskian $W(\omega, \theta)$:
\begin{equation}
W(\omega, \theta) = {a \over \tau}
.\label{wron}
\end{equation}

\noindent
Recall that the operator appearing on
the left-hand side of eqs. (\ref{acons3})
 is the form taken by $\partial^2$
acting on a
field depending on $\tau$ alone.
Furthermore, in the regime that we are now
considering, $\theta \approx \pi_b$ and
$\omega \approx \pi_c$. Hence,
eqs. (\ref{acons3}) describe the free
propagation of massive pions,
the mass being
$H^{{1 \over 2}}$, as expected. Eqs.
(\ref{acons3}) become Bessel equations
when the variable
$\sqrt{H} \tau$ is used. One has

\begin{eqnarray}
\theta & \approx & {D \over
{(\sqrt{H} \tau)^{{1 \over 2}}}}
\cos{(\sqrt{H} \tau + \zeta)} \nonumber \\
\omega & \approx & {a \over
{D(\sqrt{H} \tau)^{{1 \over 2}}}}
\sin{(\sqrt{H} \tau + \zeta)},
\label{sol}
\end{eqnarray}

\noindent
for $\sqrt{H} \tau \gg 1$, where $D$
and $\zeta$ are integration
constants. Notice, that the amplitudes
and phases of $\theta$ and $\omega$
are related by the constraint (\ref{wron}).
Eqs. (\ref{sol}) hold asymptotically,
independently of the values taken
by the constants $a$ and $b$.
\par
In order to determine the constant $D$
one needs to match the two extreme
regimes discussed so far, which is
not an easy task if attempted in full
generality but which simplifies
considerably when $a \ll b$.
\par
By using the approximation already
 discussed in which one replaces $r$ by its
expectation value, and $r r^\prime$ by
zero, and assuming that $\omega$
remains small enough, we can transform
eq. (\ref{acons0}) for $\pi_b$ into

\begin{equation}
\theta'' + {1 \over \tau} \theta' +
H \sin{\theta} =0 .
\label{theta}
\end{equation}

\noindent
This is the equation of a pendulum
subject to a time dependent friction. The
asymptotic solution,  for $\tau \to \infty$,
has already been written in eq.
(\ref{sol}) and describes an oscillatory
motion. The initial value of $\theta$
is not very important. However, for large
$b$, the derivative $\theta' =
b/\tau$ remains large, generically,
at the time when the expectation value of
$r$ becomes close to unity. As long as the angular
velocity  is large enough, the motion is
circular. Our purpose, now, is to discuss
the transition from the circular to
oscillatory motion of this  simple
dynamical system. In particular, we
shall determine the time when the
transition takes place and the amplitude of
final oscillations (the parameter
$D$ in eqs. (\ref{sol})). \par Eq.
(\ref{theta})
implies that

\begin{equation}
{d \over {d\tau}} E(\tau) \equiv {d \over {d\tau}}
[{1 \over 2} (\theta')^2 + H(1 - \cos{\theta})] =
 -{1 \over \tau }(\theta')^2 < 0 .
 \label{energy}
\end{equation}

\noindent
Thus the ``energy" $E(\tau)$ of the
pendulum is a strictly decreasing
function of time. As long as the total
 energy is much larger than the potential
energy $H(1 - \cos{\theta})$ , the
motion is  circular. The transition occurs,
when $E(\tau)$ ``hits" the potential. Let
us solve eq. (\ref{theta}),  treating
this potential as a perturbation. One finds

\begin{equation}
\theta \approx b \ln{(\tau/\tau_0)},
\label{est}
\end{equation}

\noindent
and

\begin{equation}
\theta'\approx {b \over \tau} +
 H {\tau \over b} \cos{[b
\ln{(\tau/\tau_0)}]}.
\label{theta'}
\end{equation}

\noindent
 Since the
slope of $E(\tau)$ decreases
rapidly with $\tau$, for large $b$
the potential is ``hit" near its top at time
$\tau$ satisfying $\cos{[b\ln{(\tau/\tau_0)]}}
\approx - 1$. The transition
time is estimated by setting $\theta' = 0$
and is found to be

\begin{equation}
\tau_{trans} \approx {b \over \sqrt{H}} , \; \; \; b \gg 1.
\label{trans}
\end{equation}

\noindent
As one might have expected,
the transition occurs when
the naively estimated kinetic energy
becomes comparable to the potential
energy. The amplitude of oscillation
that results is, of course bounded by
$\pi$. Actually, it is close to $\pi$ just after
the transition, precisely because
 the potential is ``hit" near its top.
Linearizing the potential one gets
the solution (\ref{sol}). Keeping in
mind the remark just made about
the amplitude, and using (\ref{trans}) one finds

\begin{equation}
D \approx {\rm const.} \sqrt{b} , \; \; \; b \gg 1,
\label{D}
\end{equation}

\noindent
where the constant is of order unity.
The amplitude of the pion field is
$\propto \sqrt{b}$ and, consequently,
the height of the
rapidity plateau is $\propto b$.
This is similar to the result found in
\cite{bk} in the framework of the non-linear
$\sigma$ model. As was the case there, the
 parameter $b$ ($\approx\kappa$
when $a\ll b$) is a measure of the energy
 released in the decay of the
condensate.

\section{DISCUSSION}
We have identified several
stages in the evolution
of the chiral condensate. First there
is a short phase in which
angular motion and radial motion
are strongly coupled. This phase lasts for
a time $\sim b/\sqrt{2\lambda}$.
Next, the fictitious particle rotates slowly,
while oscillating rapidly about the
equilibrium radial position $r\approx 1$.
Because of the expansion of the
physical system, the energy of the fictitious
particle is damped. At some point
there is  a transition from
circular to oscillatory motion, which
actually corresponds  to free
propagation of final state pions. The
transition takes place when the mass of
pion excitation can no longer be
neglected, and it occurs typically when
$\tau \sim b/\sqrt{H}$. \par
The pion
field has a random orientation in
isospace and therefore the fraction $f$
of neutral pions is distributed according to the law

\begin{equation}
dP(f) = {1 \over {2\sqrt{f}}} df
\label{f}
\end{equation}

\noindent
written first explicitly in ref. \cite{bk}
(but obtained earlier, and
independently,  by Bjorken \cite{bjf}).
At all  times the
system is subject to ``friction". The
latter is of purely kinematical origin
and reflects the boost invariance imposed
on our solution. There is no
true energy dissipation. Simply, the
expansion causes the field energy  in
a covolume to decrease.
\par
 The two time scales $\tau_1=
(2\lambda)^{-{1\over
2}}$ and $\tau_2 = (H)^{-{1 \over 2}}$, if
estimated using the
phenomenological values of the pion and
$\sigma$ masses, differ only by
a factor of 4. This is presumably not
sufficient to guarantee a clean separation
of regimes, especially when $b$ is not large
enough (and may be related to difficulties in
producing large domains of misaligned vacuum
reported in ref. \cite{ggp}). Although it seems
that the picture becomes sharper
when the condensate carries more energy,
there is a price to be paid for that: as
will be shown in a moment, the probability
that $b$ is large falls exponentially.
\par
Let the fields ($\vec{\pi}, \sigma$)
and their (proper) time
derivatives be Gaussian random variables
with variances $\sigma_1$ and $\sigma_2$,
respectively, at some initial time
$\tau_0$ belonging to the chiral
symmetric regime (cf. ref. \cite{rw}).
With this assumption let us calculate the
probability $P(\kappa, \omega)$ to
find a given value of $\kappa =
\sqrt{a^2 +
b^2}$ and  $\omega = \arctan{(a/b)}$.
Using the polar parametrization of the
fields in the randomly oriented hyperplane
$\perp \vec{a}$, together with the
constraint $\kappa = \tau_0 r^2\theta'$
(cf. (\ref{rot})), one finds, after a
straightforward calculation:

\begin{equation}
{{d^2} \over {d\kappa d\omega}}
P(\kappa, \omega)=
{\kappa \over {2(\pi \sigma_1
\sigma_2 \tau_0)^2}}
K_1 \left( {\kappa \over {\sigma_1
\sigma_2 \tau_0}} \right) ,
\label{prob}
\end{equation}

\noindent
where $K_1(z)$ denotes the second
modified Bessel function,
$K_1(z) \sim \sqrt{\pi/2z} e^{-z}$
for $z \to \infty$. The probability distribution
(\ref{prob}) falls exponentially at
large $\kappa$. This result is unlikely
to be of much phenomenological
relevance,  since it is obtained in a
$1+1$ dimensional model. The reduction
to $1+1$ dimensions means full
translational symmetry in transverse
coordinates and, in particular,
coherence over large transverse distance.
The probability that such coherence
does occur cannot be estimated within the
framework of the model. Eq.
(\ref{prob}) represents nevertheless a
significant conceptual progress with
respect to the discussion of refs.
\cite{ans} -\cite{bk}:  the present
approach, which has been partly inspired by ref.
\cite{rw}, enables one to determine
the absolute probability of observing a
disoriented chiral condensate with
given global characteristics.
\par
The initial randomness of the fields
($\vec{\pi},\sigma$), and their
derivatives, does not necessarily reflect
thermalization occuring at the early stage of
the collision  process. Actually, the
most interesting aspect of the quench
model put forward, in this context, by
Rajagopal and Wilczek is the out of
equilibrium evolution of the system.
In discussing this evolution one does
not need to talk about temperature at all.
Since the seminal work of Landau
\cite{lan}, the study of complicated nuclear
collisions uses the concept of
(local) thermal equilibrium as the central piece of
the paradigm. If we were to assume
that local thermal equilibrium is
maintained throughout the time
development, we would be led to use
hydrodynamical evolution equations.
The resulting evolution would be very
different from that obtained here.
It is unclear, at this moment, whether
the disoriented chiral condensate
can be really produced in high energy
collisions. However, the theoretical
investigation of this as yet
undiscovered phenomenon has already
turned out to be useful, since it has been a good
pretext for trying to figure out non-equilibrium aspects
of high-energy nuclear collisions.
\par\noindent
\begin{center}
{\bf ACKNOWLEDGEMENTS}
\end{center}
\par
An interesting conversation with A. Bia\l as and a
helpful suggestion from J. Ginibre are acknowledged.
\pagebreak

\end{document}